\documentclass[a4paper,twoside]{article}

\usepackage{epsfig}
\usepackage{subcaption}
\usepackage{calc}
\usepackage{amssymb}
\usepackage{amstext}
\usepackage{amsmath}
\usepackage{amsthm}
\usepackage{multicol}
\usepackage{pslatex}
\usepackage{apalike}
\usepackage[bottom]{footmisc}
\usepackage{xcolor}
\usepackage{url}

\usepackage{SCITEPRESS}

\begin{document}
\title{Interoperability-oriented Quality Assessment for Czech Open Data}


\author{\authorname{Dasa Kusnirakova\sup{1}, Mouzhi Ge\sup{2}, Leonard Walletzky\sup{1}, and
Barbora Buhnova\sup{1}
}
\affiliation{\sup{1}
Faculty of Informatics, Masaryk University, Brno, Czech Republic}
\affiliation{\sup{2}Deggendorf Institute of Technology, Deggendorf,  Germany}
\email{\{kusnirakova, wallet, buhnova\}@mail.muni.cz, mouzhi.ge@th-deg.de}
}

%
%

%
\keywords{Open data, data quality, data interoperability, evaluation framework}

\abstract{With the rapid increase of published open datasets, it is crucial to support the open data progress in smart cities while considering the open data quality. 
In the Czech Republic, and its \textit{National Open Data Catalogue} (NODC), the open datasets are usually evaluated based on their metadata only, while leaving the content and the adherence to the recommended data structure to the sole responsibility of the data providers. The interoperability of open datasets  remains unknown. This paper therefore aims to propose a novel content-aware quality evaluation framework that assesses the quality of open datasets based on five data quality dimensions. With the proposed framework, we provide a fundamental view on the interoperability-oriented data quality of Czech open datasets, which are published in NODC. Our evaluations find that domain-specific open data quality assessments are able to detect data quality issues beyond traditional heuristics used for determining Czech open data quality, increase their interoperability, and thus increase their potential to bring value for the society. The findings of this research are beneficial not only for the case of the Czech Republic, but also can be applied in other countries that intend to enhance their open data quality evaluation processes.
}

\onecolumn \maketitle \normalsize \setcounter{footnote}{0} \vfill

\section{Introduction}
With a broader adoption of the open data paradigm worldwide and the increasing number of published datasets, the focus of dealing with the data has been shifted and data quality has become a major concern in organizations \cite{intro:quality-over-quantity}. The uncertain data quality is becoming critical since low-quality datasets have very limited capacity of creating added value, sometimes may even harm the applications and services \cite{Ge20}. 

One of the key issues is the interoperability of open datasets, as the concept of open data is based on the rationale of reusability and interconnection with other data \cite{intro:opendata}. As the data comes from different publishers, its structure or individual values are often incompatible with each other \cite{GeC0P19}. Such diversity of datasets, e.g., in terms of data formats or vocabularies used, then significantly increases the processing effort for further data usage \cite{intro:odi-interoperability}, or may even make data merging completely impossible.


The aim of this paper therefore is to propose a novel quality evaluation model that measures the quality of open datasets across five major data quality characteristics. With taking datasets' interoperability as a priority, the evaluation framework focuses on the interoperability of the datasets, taking the content and the data structure into consideration. Besides that, we present the insights on interoperability-oriented data quality assessment for Czech datasets in the tourism domain, which are published in Czech \textit{National Open Data Catalogue} (NODC). Finally, we argue that domain-specific and interoperability-oriented open data quality assessment is capable of identifying multiple serious data quality concerns in addition to the usual techniques used to assess Czech open data quality. 

\section{Related Work}
\label{sec:relwork}

In recent years, there has been research progress on addressing the issue of open data quality. For example, in \cite{relwork:5star}, the author proposed \textit{5-Star Open Data Rating System}. The data subjected to quality analysis is awarded a certain number of stars according to defined quality requirements. Even though this tool is widely used within Europe and is being promoted by the European Data Portal \cite{relwork:goldbook}, its results may be misleading. The evaluation focuses only on a subset of data
quality dimensions (mainly legal and technical aspects), and therefore may not reflect all the user demands on data quality.

The majority of the published papers investigates open data portals. In \cite{relwork:ubaldi} and \cite{relwork:viscusi} the authors conducted thorough studies for evaluating the quality of Open Government Data. The authors created a series of metrics determining data quality by its availability, demand, re-use, format or timeliness. However, likewise in the previous work, none of the proposed metrics considers the datasets' content; all metrics operate only on the dataset or portal level. 

The intention to assess the quality of open data \textit{inside} data files is not completely new. One of the first evaluation metrics operating on interoperability (e.g. currentness or completeness) was introduced in \cite{relwork:vetro}. Even though the paper proposes quality dimensions at most granular level of measurement, the framework lacks syntactic and semantic assessments of the examined data. A more general approach in the form of an executable evaluation model enabling custom definition of data quality specifications was suggested by \cite{relwork:nikiforova}. This approach is, however, subject to the precise and correct specification of the data quality requirements and entails a significant amount of manual intervention.

\section{Open Formal Standards}
\label{sec:ofn}
One of the standardization techniques introduced in the Czech Republic aiming to ensure interoperability of open datasets is called Open Formal Standards (OFSs). OFSs are technical guidelines created for selected domains, developed within a collaborative decision-making process coordinated by the Czech Ministry of the Interior. They aim to simplify data usage and ensure data interoperability, even when various data providers provide the same kind of data. Full interoperability is~ensured in technical, syntactic as well as semantic dimension \cite{ofn:data-fond}. 

The assurance of interoperability represents the main reason why these standards should be adhered to if a publisher publishes data relevant to what OFSs are modeling. Apart from that, OFSs are binding on open data publishers according to Act No. 106/1999 Coll. on Free Access to Information.

\begin{figure}
  \includegraphics[width=\linewidth]{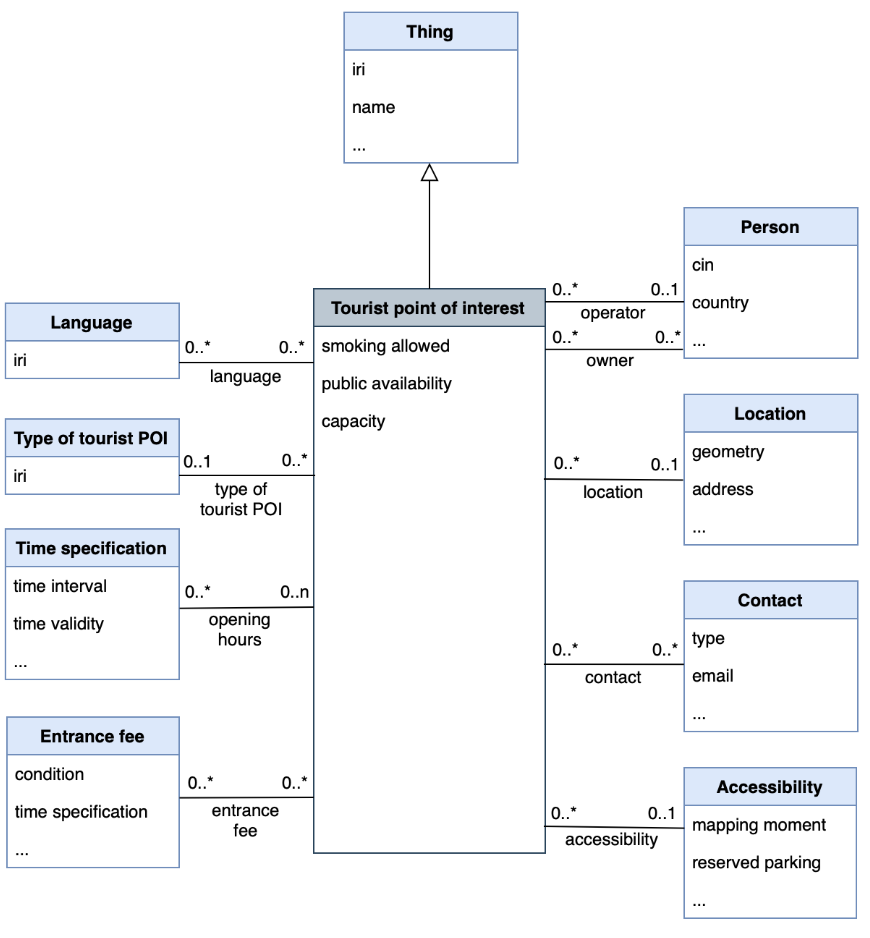}
  \caption{An example of OFS scheme for objects regarding \textit{Tourist points of interest} \cite{ofn:format}. Translated to English.}
  \label{fig:ofn}
\end{figure}

Each OFS begins with a description of essential terms for a given dataset which unifies the semantics of data; that is~how the data is~understood. The terms are represented in the form of~a~conceptual scheme, which models the terms as classes, their properties, and the relationships between them \cite{ofn:content}. The scheme is also illustrated graphically for easier understanding as displayed in Figure \ref{fig:ofn}. This format is uniform for all issued OFSs. 

One of the key aspects of OFSs is that classes that appear in several OFSs, such as \textit{Person}, \textit{Contact}, or \textit{Location}, are specified in a single place, so-called shared specifications. Shared specifications ensure compatibility between the same entities in different datasets and thus facilitate data processing according to various OFSs. For example, information about admission to a concert provided in a dataset adhering to OFS for \textit{Events} is represented in the same way as admission to a castle published in a dataset regarding \textit{Tourist points of interest}.

\subsection{Current state of OFS adoption}
The Czech National Open Data Catalogue does not check individual datasets' quality; NODC only works with the metadata of the corresponding data files \cite{relwork:klimek}. Even though the Ministry of the Interior supports data providers in improving the quality of their data through various tools, such as regular evaluation of datasets' metadata or providing most common bad practice examples \cite{tools:bad-practice}, interoperability-oriented data quality of individual datasets is solely the responsibility of data providers \cite{ofn:nkod-analysis}.

Even though OFSs show a great deal of potential and are legally binding on data publishers, their actual application in practice is unsatisfactory. We checked the current compliance with OFSs regarding \textit{Tourist points of interest} manually with the usage of Open Data National Catalogue API, where we searched for the datasets having the \textit{Tourist points of interest} standard's identifier given as a value of any attributes in the dataset's metadata, as defined by the OFS. The results were surprising. None of the published datasets regarding this topic has adhered to this standard yet. These findings prompted us to research the subject further.

\section{Quality Assessment Framework for Czech Open Data}
\label{sec:framework}

To design the evaluation framework, we applied the method by \cite{fw:beyond-accuracy}. The proposed data quality framework was designed with comprehension of~what characteristics are relevant for Czech open data, while maintaining the ease of comprehensibility of the results and metrics even for the general public. The general features of the evaluation framework are described below.

\subsection{Score calculation}
Each dimension of the proposed framework awards a certain number of points to a dataset based on its quality within the examined aspect. The given score ranges from 0 to 100 points, while the higher is~the~score, the higher is~the~dataset's quality. The calculated scores for individual dimensions remain separate and are not combined into one final score. Their purpose is to highlight the dataset's strengths and flaws in terms of~interoperability and adherence to the particular OFS.

\subsection{Features}
\label{subsec:features}
Each OFS typically contains a minimal example specifying minimum requirements on modeled entities. If a dataset contains less information than provided in the minimal sample, the data is most likely meaningless, and no one will be able to use it \cite{ofn:general}. Therefore, these minimal requirements, denoted as features, are considered \textit{mandatory} by the proposed evaluation framework, even though all entities specified in the OFS schema are optional, according to the official documentation \cite{ofn:docs}. 

Besides the minimal example, OFSs provide a more complex example, too. Such a complex model portrays elements that should be filled in to provide the user with the most accurate picture of the modeled domain. The additional features are considered \textit{optional} by the proposed framework. 

\subsection{Feature weight}
Because each feature has different importance, their impact on the resulting score of the examined dimension should also vary. The absence of a mandatory feature (e.g., the location of a tourist destination) causes much more significant issues in the data processing phase than absence of an optional one (e.g., the languages spoken at the destination). Such importance of~an~individual feature is denoted as feature weight.

Feature weight is determined by its type. The weight of mandatory features is significantly higher since they carry the most critical information; their absence results in a considerable drop in the dataset's quality. In~particular, the proposed framework determines that the weight of one mandatory feature is equal to the sum of the weights of all optional features. Because the model operates on a scale ranging from 0 to 100, the weight calculation is performed as follows:

\begin{equation}
\label{eqn:weight-calc-1}
\Sigma \text{ } w_{mf} + \Sigma \text{ } w_{of} = 100 
\end{equation}

\begin{equation}
\label{eqn:weight-calc-2}
w_{mf} = \frac{100}{N_{mf} + 1} 
\end{equation}

\begin{equation}
\label{eqn:weight-calc-3}
w_{of} = \frac{w_{mf}}{N_{of}}
\end{equation}

\noindent where $w$ denotes a feature weight, $N$ represents the number of~features, $mf$ stands for mandatory feature and $of$ is an optional feature. 

\section{Data Quality Dimensions}
\label{sec:dimensions}

Throughout the development process, we identified ten quality dimensions, from which those five selected seemed to be the most clear candidates in terms of assessing datasets' a) interoperability, b) adherence to the rules defined by OFSs, and c) essential quality aspects, which are currently missing in Czech NODC.

\subsection{File format}
Prior to combining diverse data sources, the data format is one of~the key aspects that need to be considered \cite{dim:data-profiling}. Different data formats require different ways of data processing. Moreover, each data format places other requirements on the data structure, which in principle worsens data interoperability.

Data processing may be automated for many common formats. Nevertheless, the process gets more complicated if various data formats are used for the same data representation. It may even require manual assistance in case of inconsistencies caused by the nature of~particular file formats, especially when also the data structure needs to~be~changed \cite{dim:data-wrangling}. 

\paragraph{Formal requirements}
According to the OFS specification, open datasets need to be published in JSON format \cite{ofn:format}. However, Czech datasets published in NODC use other formats, too, such as CSV, XLSX, XML, or special formats from the JSON family, which causes issues for data to interoperate easily.

\paragraph{Score calculation}
The \textit{File format} dimension awards dataset a score based on its data format, and the score is given according to the conversion rules displayed in Table \ref{tab:file-format}. The individual levels of the proposed table have been designed based on the format's similarities to the required JSON format. The levels can also be understood to represent the ease of transformation from the particular file format to JSON. 

\begin{table}[h]
\centering 
{\scriptsize
\caption{Conversion table for awarding scores based on the dataset's file format}
\label{tab:file-format}
\begin{tabular}{|c|l|}
\hline
\textbf{Score} & \textbf{File format}  \\ \hline
100            & JSON, JSON-LD, GEOJSON \\ 
75             & XML, GML, KML, RDF     \\ 
50             & CSV                    \\ 
25             & XLS, XLSX              \\ 
0              & PDF, TXT               \\ \hline
\end{tabular} }
\end{table}

The highest score is given to file formats from the JSON family, as defined by the OFS. The second-best score is awarded to datasets with file format from the XML family. XML documents allow hierarchy in the same way as JSON and besides that, the data format can be simply converted to JSON in most cases by various online tools. Datasets with a tabular structure, such as CSV files, are awarded 50 points. This file format does not allow hierarchical structure by nature and therefore is not suitable for complex data. The second-lowest score is given to datasets with tabular structure that are published in a proprietary file format like XLS, where a special software may be required to read data properly, which contradicts the general concept of open data. No points are given to files published in a format that does not guarantee any structure, e.g. TXT or PDF.

\subsection{Schema accuracy}
\textit{Schema accuracy} refers to the syntactic accuracy of features' names compared with the naming convention defined by OFS. The focus on~the~naming of modeled entities is an essential part of quality assessment with an emphasis on interoperabilty, as two entities cannot be merged automatically if their names are different, even if they semantically represent the same object. A real example of incorrectly named features can be, for instance, using English feature names such as \textit{location} instead of the Czech word \textit{umístění}.

\paragraph{Formal requirements}
Requirements placed by OFS require a certain number of mandatory and optional features, while specifying their correct naming. Except for the unique key \textit{@context}, specified by JSON-LD format~\cite{dim:jsonld} and used for interlinking the dataset with the corresponding standard against which a dataset structure is valid, all feature names are in~the Czech language. There is always only one correct feature name.

\paragraph{Score calculation}
The dataset's score in terms of schema accuracy is~calculated based on the correctness of individual feature names. The correctness of the naming of mandatory and optional features impacts the score in various ways, as presented in the Equation \ref{eqn:sch-acc} below.
 
\begin{equation}
\label{eqn:sch-acc}
score = \sum_{f=1}^{F} w_{f} n_{f} \text{  }
\begin{cases}
    n=1,& \text{if } n \in N\\
    n=0,              & \text{otherwise}
\end{cases}
\end{equation}

\noindent where $f$ denotes a feature specified by the standard, $w_{f}$ is the weight of a feature, and $n_{f}$ represents the name of a feature. A set of both mandatory and optional features specified by OFS are denoted by~$F$, while $N$ represents the set of all feature names contained in the examined dataset.

\subsection{Schema completeness}
\textit{Schema completeness} is dedicated to checking the semantic correctness of the information carried by individual features. In other words, the dimension focuses on the features' content, regardless of the correctness of their naming. Data types used are also ignored in this case. Before the actual score calculation, each examined dataset needed to be manually adjusted - the original features' names have to be changed to match the words specified by OFS in case the information carried by the feature semantically matches the standard.
 
\paragraph{Formal requirements} 
The standard defines a set of mandatory and optional features understood as pieces of information that a dataset should contain, as discussed in section \ref{subsec:features}. These features represent information necessary for providing the user with the most accurate picture of the modeled domain. Besides that, the information referring to the used standard is also required. 

\paragraph{Score calculation} The score for schema completeness is determined in a similar way as for the schema accuracy dimension. However, the dimension's focus is different. The final score is based on the presence or absence of expected information contained in a dataset, as given by Equation \ref{eqn:sch-comp}.

\begin{equation}
\label{eqn:sch-comp}
score = \sum_{f=1}^{F} w_{f} i_{f} \text{  }
\begin{cases}
    i=1,& \text{if } i \in I\\
    i=0,              & \text{otherwise}
\end{cases}
\end{equation}

\noindent where $f$ denotes a feature specified by the standard, $w_{f}$ is the weight of a feature, and $i_{f}$ represents the information carried by a feature. A set of both mandatory and optional features specified by OFS is denoted by $F$, while $I$ represents the set of the expected information contained in the dataset according to the standard.

\subsection{Data type consistency}
\label{dim:type-cons}
The quality of inputs determines the ability to fuse digital data and create relevant information \cite{dim:fusion}. Because the same kind of data can be produced in various formats and using different data types for the same data requires time-consuming data cleansing, it is suitable to monitor the aspect of data consistency. In particular, this dimension focuses on the consistency of data types used for the same data representation within a feature.

The implementation of the proposed evaluation model can distinguish five primary data types, namely \textit{integer}, \textit{float}, \textit{bool}, \textit{string}, and \textit{null}. On~top of that, we have decided to extend the type recognition by five custom data types, which were selected based on the analysis of the actual values provided in the examined datasets. The list of the custom data types is as follows:

\begin{itemize}
    \item \textit{URL}: recognized by a function for URL recognition contained in~a~python library \textit{validators},
    \item \textit{E-mail}: recognized by a function for e-mail recognition contained in a python library \textit{validators},
    \item \textit{Address}: identified by a regex string searching for an address written in Czech format consisting of \textit{street name} and \textit{number} in~given order,
    \item \textit{Point}: recognized by a pattern, where geographical coordinates are wrapped by the \textit{POINT} label,
    \item \textit{Phone number}: identified by a regex string searching for a phone number, written in Czech or international form.
\end{itemize}

\paragraph{Formal requirements} All the values within one feature should be represented by the same data type. 

\paragraph{Score calculation} The score is affected by the number of used data types within one feature and its weight. Since the \textit{Data type consistency} dimension focuses on the values themselves rather than on their comparison against the standard's schema, there is a need for a minor adjustment in the definition of mandatory and optional features from the definition presented in section \ref{subsec:features}:

\begin{itemize}
    \item \textit{Mandatory features}: Features defined as mandatory by the standard, which are at the same time contained in the examined dataset. 
    
    \item \textit{Optional features}: All other features which are included in the examined dataset.
\end{itemize}

In other words, the feature weights are derived from the total number of provided features within a dataset, not solely from the features defined by OFS. Suppose a dataset containing two mandatory features out of seven defined by the standard, and zero optional features. Then the weight of each feature is 50. Compulsory features which are missing have no effect on the score calculated for this dimension.

This means that even if a dataset is missing some mandatory features and does not include any optional ones, it can still score a maximum of 100 points in terms of data consistency. Naturally, all values within each feature must be represented by the same data type in such a case, otherwise the score gets lower. 

Multiple data types used within a single feature are stringently punished. The formula for score calculation is provided in Equation \ref{eqn:type-cons}.

\begin{equation}
\label{eqn:type-cons}
score = \sum_{f=1}^{F} w_{f} \text{ } \frac{1}{t_{f}}
\end{equation}

\noindent where, $f$ denotes a feature provided in the dataset, $w_{f}$~is~the feature weight, and $t_{f}$ represents the number of data types used within a~feature. A set of all features provided in the dataset is denoted by~$F$.

\subsection{Data completeness}
\label{dim:data-comp}

The amount of data we collect is growing. However, not all of the data is complete, and some information can be missing. Missing values are typically denoted with a null value, a specific mark indicating that a~value is absent or undefined \cite{dim:null-codd}, but empty strings are also widely~used. 
As stated in the literature, there are two main reasons for providing incomplete data \cite{dim:null-codd}. The information may be unknown to the data source, or it refers to a property that is not relevant to the particular object. In any case, null values represent a severe quality issue. They bring confusion to further data processing, and may lead to wrong data interpretations, or completely restrict data to be usable. Therefore, the data processor needs to be aware of them in order to choose the right strategy for their correct handling. 

\paragraph{Formal requirements} All values within a feature should ideally be complete; that is different from null and empty objects.

\paragraph{Score calculation} 
The data completeness dimension measures the weighted ratio of non-null values. By ratio, we mean the division of~non-null values by all values provided within the examined feature. Both values of data type null and empty strings are understood as~null values in the proposed model. 

As for the weights, their calculation process is the same as described in section \ref{dim:type-cons}. The score for the \textit{Data completeness} dimension is calculated as stated in Equation \ref{eqn:data-comp}.

\begin{equation}
\label{eqn:data-comp}
score = \sum_{f=1}^{F} w_{f} \text{} \frac{n_{f}}{v_{f}} 
\end{equation}

\noindent where $w_{f}$ is the weight of a feature, $n_{f}$ represents the number of non-null values within a feature, and $v_{f}$ denotes the number of all values within the examined feature. A set of both mandatory and optional features provided in the dataset is marked as $F$. 

\section{Evaluation}
\label{sec:application}
We have applied the evaluation framework introduced in section \ref{sec:dimensions} to measure the data quality of Czech open datasets on \textit{Tourist points of interest} domain, with focus on their interoperability and their adherence to the standards developed by the Czech Ministry of the Interior. As the data quality of datasets published in NODC is currently monitored only on the metadata level, their interoperability quality of data remains unknown. 

Overall, we collected 14 datasets from six different municipalities, all published in NODC. Because the quality of datasets from the same provider was essentially the same, we decided to analyze only one dataset from each provider so that the results were unbiased. Therefore, the evaluation concerns six datasets from six distinct municipalities, which are listed in Table \ref{tab:datasets}.

Each of the selected datasets provides a different number of~records and features. The overview is listed in Table \ref{tab:data-stats}. While the dataset from Huntířov municipality contains 10 records, Brno offers a slightly more rich collection of 350 tourist destinations. As~for the features, both Praha and Brno datasets capture 9 pieces of information regarding each record. The largest number of 71 features offers datasets from Děčín. However, it is necessary to note that many features in this dataset do not contain any meaningful value other than null.

\begin{table}[h]
\centering {\scriptsize
\caption{List of municipalities providing data regarding \textit{Tourist points of interest} and selected datasets for analysis}
\label{tab:datasets}
\begin{tabular}{|l|c|c|c|}
\hline
\multicolumn{1}{|c|}{\textbf{Municipality}} & \textbf{\begin{tabular}[c]{@{}c@{}}Number of\\ relevant\\ datasets\end{tabular}} & \textbf{\begin{tabular}[c]{@{}c@{}}Selected dataset\end{tabular}} \\ \hline
Brno & 2 & Turistická místa \\ \hline 
Děčín & 1 & \begin{tabular}[c]{@{}c@{}}Seznam bodů\\ zájmů (POI)\end{tabular} \\ \hline
Hradec Králové & 8 & Zámky \\ \hline
Huntířov & 1 & Turistické cíle \\ \hline
Ostrava & 1 & Turistické cíle \\ \hline
Praha & 1 & \begin{tabular}[c]{@{}c@{}}Významné \\vyhlídkové body\end{tabular} \\ \hline
\end{tabular} }%
\end{table}

\begin{table}[h]
\centering {\scriptsize
\caption{Overview of the number of records and features provided by the selected datasets}
\label{tab:data-stats}
\begin{tabular}{|l|c|c|c|}
\hline
\multicolumn{1}{|c|}{\textbf{Municipality}} & \textbf{\begin{tabular}[c]{@{}c@{}}Number\\ of records\end{tabular}} & \textbf{\begin{tabular}[c]{@{}c@{}}Number\\ of features\end{tabular}} \\ \hline
Brno & 350 & 15 \\ \hline
Děčín & 254 & 71 \\ \hline
Hradec Králové & 33 & 9 \\ \hline
Huntířov & 10 & 18 \\ \hline
Ostrava & 66 & 18 \\ \hline
Praha & 323 & 9 \\ \hline
\end{tabular} }
\end{table}

\begin{table*}[t] \centering {\scriptsize
\caption{Complete results of earned scores in five monitored dimensions for individual municipalities}
\label{tab:results}
\begin{tabular}{|l|c|c|c|c|c|}
\hline
\multicolumn{1}{|c|}{\textbf{Municipality}} & \textbf{\begin{tabular}[c]{@{}c@{}}File \\ format\end{tabular}} & \textbf{\begin{tabular}[c]{@{}c@{}}Schema \\ accuracy\end{tabular}} & \textbf{\begin{tabular}[c]{@{}c@{}}Schema \\ completeness\end{tabular}} & \textbf{\begin{tabular}[c]{@{}c@{}}Data type \\ consistency\end{tabular}} & \textbf{\begin{tabular}[c]{@{}c@{}}Data\\ completeness\end{tabular}} \\ \hline
Brno & 100 & 0 & 65.38 & 53.49 & 64.46 \\ \hline
Děčín & 25 & 0 & 53.85 & 58.03 & 75.99 \\ \hline
Hradec Králové & 100 & 0.96 & 38.46 & 97.94 & 100 \\ \hline
Huntířov & 50 & 0 & 52.88 & 70.97 & 76.3 \\ \hline
Ostrava & 50 & 0 & 52.88 & 65.73 & 81.03 \\ \hline
Praha & 100 & 0 & 25 & 78.56 & 100 \\ \hline
\end{tabular} }
\end{table*}

\subsection{Results}
\begin{figure}
  \includegraphics[width=211px]{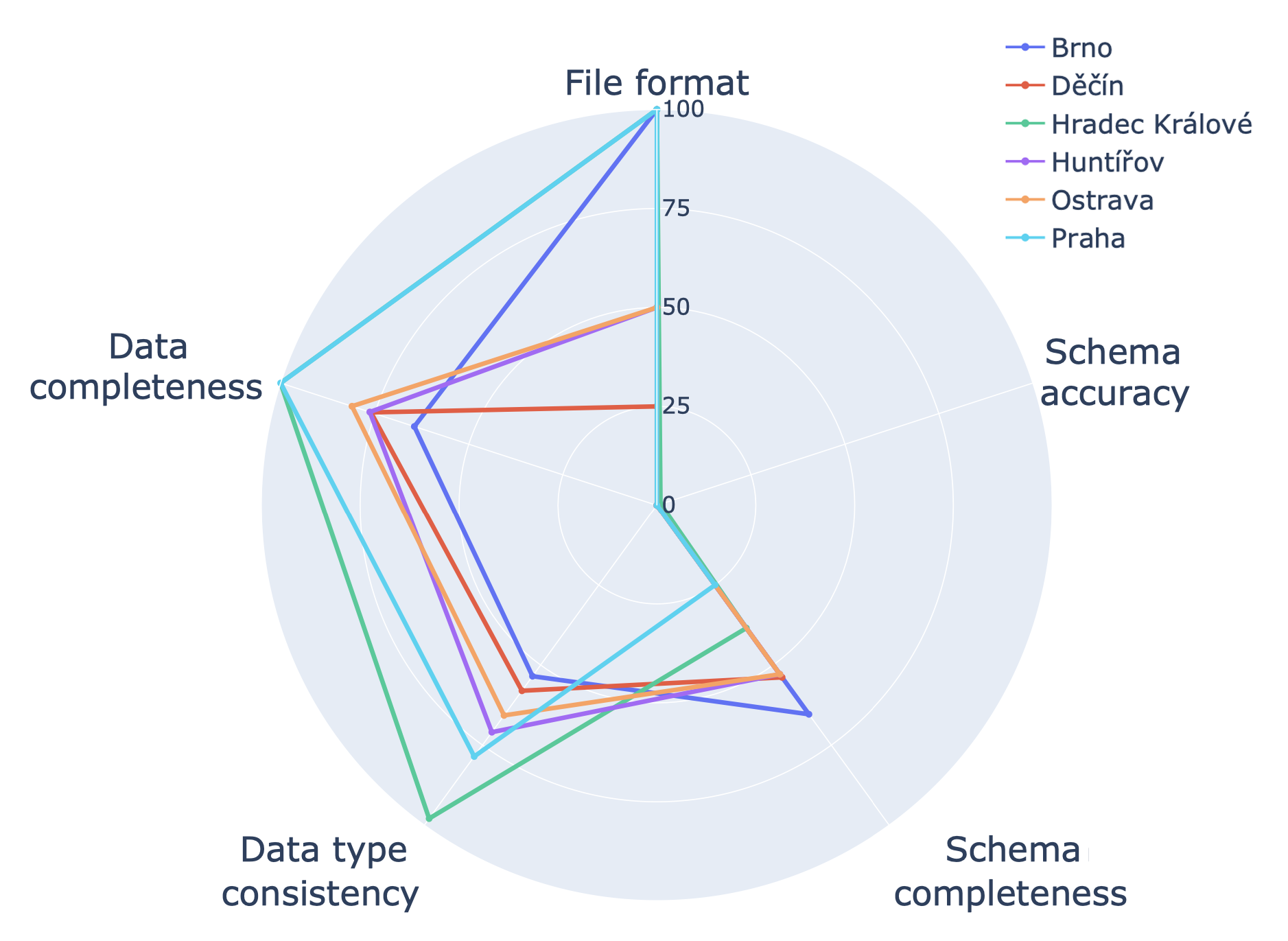}
  \caption{Comparison of the achieved scores in each data quality dimension for individual municipalities}
  \label{graph:radial-one}
\end{figure}

We analyzed each dataset in terms of the five proposed data quality aspects. In each dimension, a dataset could get a score ranging from 0~up~to~the~maximum of 100 points. The full record of achieved scores in each data quality dimension are provided in~Table \ref{tab:results}. A better overview of the results is then visualized by~a~radar chart in Figure \ref{graph:radial-one}.  

\subsection{Result analysis}
The findings are diverse; while datasets generally achieve a somewhat satisfactory score in some dimensions, such as \textit{Data completeness}, other aspects, especially \textit{Schema accuracy}, prove significant deficiencies for all datasets. In the following paragraphs, we focus on the dimensions with the most shortcomings identified, and provide an in-depth analysis of the results.

\paragraph{Schema accuracy}
There are multiple reasons why all datasets achieve unsatisfactory results in \textit{Schema accuracy} dimension. After a thorough review of the feature names, we identified the following causes:

\begin{itemize}
 \item two datasets (Brno, Hradec Králové) use English feature names,
 
 \item two datasets (Huntířov, Ostrava) use Czech feature names, but without diacritics or modified in any other way,
 
 \item two datasets (Děčín, Praha) use a combination of both English and Czech feature names, even without diacritics or modified in~any~other way.
\end{itemize}

We observed that each municipality uses its own set of unique feature names. However, certain similarities can be seen in the datasets. For example, the datasets from Ostrava and Huntířov share 10 out of 18 feature names, even though they differ from feature names required by the standard.

\paragraph{Schema completeness}
The results showed that despite inaccurate feature naming, the examined datasets do contain the information required by the standard, at least to some extent.

The fact that the datasets are often provided as a list of places focused simply on their location was recognized as the primary source of the observed difficulties in terms of schema completeness. This is especially true for datasets published in GEOJSON format, designed specifically for this purpose. But even a dataset that might be of sufficient quality for one purpose may not be suitable for another \cite{intro:quality-over-quantity}. Therefore, it would be desirable to append the missing information in~order to increase the possibilities of employing this data in the tourism industry.

\section{Conclusion}
\label{sec:conclusion}
In this paper, we have proposed an interoperability-oriented quality assessment  framework that consists of five data quality dimensions. The selection of the data quality dimentions is based on the interoperability of datasets and adherence to specified standards, known as OFSs. In order to evaluate the applicability of the proposed framework, we have assessed the datasets' quality on tourist points of interest. The datasets were downloaded from the Czech Open Data National Catalogue, and the evaluation has revealed the quality issues in the Czech Open Data National Catalogue.

While the assessed datasets have shown moderate deficiencies in \textit{Data completeness} and \textit{Data type consistency} dimensions, \textit{Schema accuracy} results turn out to be the poorest across all the dimensions. All the datasets achieved poor results in this dimension, which indicates that the selected datasets do not adhere to the standard in terms of feature naming conventions. Besides that, most datasets model individual items merely as pure localities and not as~tourist objects as~required. As a result, open Czech datasets are practically incapable of~interoperability in their current state within the tourism context, and their potential for value co-creation is decreased.

Although this paper focuses primarily on the open data quality assessment in the Czech Republic, the findings are relevant for any country that aims to improve the evaluation processes on the quality of open datasets, as the availability of the experience from different countries is crucial in the design process. Once we understand the quality of open datasets and identify their quality flaws, we can guide data producers to provide and improve data with an impact, so that the society can make full use of the open data with interoperability.


\paragraph{Acknowledgement}
This research was supported by ERDF "CyberSecurity, CyberCrime and Critical Information Infrastructures Center of Excellence" (No. CZ.02.1.01/0.0/0.0/16\_019/0000822).

\bibliographystyle{apalike}
{\small\bibliography{references}}

\end{document}